\documentclass[prl,twocolumn,floatfix,showpacs,amsmath,amssymb,superscriptaddress]{revtex4}
\usepackage{amsmath}
\usepackage{bm}
\usepackage{graphicx}
\usepackage{color}
\usepackage{colordvi}

\begin{document}

\title{Photoelectron Angular Distributions for Two-photon Ionization of Helium
by Ultrashort Extreme Ultraviolet Free Electron Laser Pulses}

\author{R. Ma}
\affiliation{Institute of Multidisciplinary Research for Advanced
Materials, Tohoku University, Sendai 980-8577, Japan}
\affiliation{RIKEN, XFEL Project Head Office, Sayo, Hyogo
679-5148, Japan}%
\affiliation{Institute of Atomic and Molecular
Physics, Jilin University, Changchun 130012, China}

\author{K.~Motomura}
\affiliation{Institute of Multidisciplinary Research for Advanced
Materials, Tohoku University, Sendai 980-8577, Japan}%
\affiliation{RIKEN, XFEL Project Head Office, Sayo, Hyogo
679-5148, Japan}

\author{K.L.~Ishikawa}
\affiliation{Photon Science Center, Graduate School of Engineering, The University of
Tokyo, 7-3-1 Hongo, Bunkyo-ku, Tokyo 113-8656, Japan}

\author{S. Mondal}
\affiliation{Institute of Multidisciplinary Research for Advanced
Materials, Tohoku University, Sendai 980-8577, Japan}
\affiliation{RIKEN, XFEL Project Head Office, Sayo, Hyogo
679-5148, Japan}%

\author{H.~Fukuzawa}
\affiliation{Institute of Multidisciplinary Research for Advanced
Materials, Tohoku University, Sendai 980-8577, Japan}%
\affiliation{RIKEN, XFEL Project Head Office, Sayo, Hyogo
679-5148, Japan}

\author{A.~Yamada}
\affiliation{Institute of Multidisciplinary Research for Advanced
Materials, Tohoku University, Sendai 980-8577, Japan}%
\affiliation{RIKEN, XFEL Project Head Office, Sayo, Hyogo
679-5148, Japan}

\author{K.~Ueda}\email[]{ueda@tagen.tohoku.ac.jp}
\affiliation{Institute of Multidisciplinary Research for Advanced
Materials, Tohoku University, Sendai 980-8577, Japan}
\affiliation{RIKEN, XFEL Project Head Office, Sayo, Hyogo
679-5148, Japan}

\author{K.~Nagaya}
\affiliation{RIKEN, XFEL Project Head Office, Sayo, Hyogo 679-5148, Japan}%
\affiliation{Department of Physics, Kyoto University, Kyoto
606-8502, Japan}%

\author{S.~Yase}
\affiliation{RIKEN, XFEL Project Head Office, Sayo, Hyogo 679-5148, Japan}%
\affiliation{Department of Physics, Kyoto University, Kyoto
606-8502, Japan}%

\author{Y.~Mizoguchi}
\affiliation{RIKEN, XFEL Project Head Office, Sayo, Hyogo 679-5148, Japan}%
\affiliation{Department of Physics, Kyoto University, Kyoto
606-8502, Japan}%

\author{M.~Yao}
\affiliation{RIKEN, XFEL Project Head Office, Sayo, Hyogo 679-5148, Japan}%
\affiliation{Department of Physics, Kyoto University, Kyoto
606-8502, Japan}%

\author{A.~Rouz$\rm \acute{e}$e}
\affiliation{Max-Born-Institut, Max-Born Strasse 2A, D-12489 Berlin,
Germany}%
\affiliation{FOM Institute AMOLF, Science park 102, 1098 XG Amsterdam, Netherlands}

\author{A.~Hundermark}
\affiliation{Max-Born-Institut, Max-Born Strasse 2A, D-12489 Berlin,
Germany}%
\affiliation{FOM Institute AMOLF, Science park 102, 1098 XG Amsterdam, Netherlands}

\author{M.J.J.~Vrakking}
\affiliation{Max-Born-Institut, Max-Born Strasse 2A, D-12489 Berlin, Germany}%
\affiliation{FOM Institute AMOLF, Science park 102, 1098 XG Amsterdam, Netherlands}

\author{P.~Johnsson}
\affiliation{Department of Physics, Lund University, 22100 Lund,
Sweden}

\author{M.~Nagasono}
\affiliation{RIKEN, XFEL Project Head Office, Sayo, Hyogo
679-5148, Japan}

\author{K.~Tono}
\affiliation{RIKEN, XFEL Project Head Office, Sayo, Hyogo
679-5148, Japan}

\author{T.~Togashi}
\affiliation{RIKEN, XFEL Project Head Office, Sayo, Hyogo
679-5148, Japan}%
\affiliation{Japan Synchrotron Radiation Research
Institute, Sayo, Hyogo 679-5198, Japan}%

\author{Y.~Senba}
\affiliation{RIKEN, XFEL Project Head Office, Sayo, Hyogo
679-5148, Japan}%
\affiliation{Japan Synchrotron Radiation Research
Institute, Sayo, Hyogo 679-5198, Japan}%

\author{H.~Ohashi}
\affiliation{RIKEN, XFEL Project Head Office, Sayo, Hyogo
679-5148, Japan}%
\affiliation{Japan Synchrotron Radiation Research
Institute, Sayo, Hyogo 679-5198, Japan}%

\author{M.~Yabashi}
\affiliation{RIKEN, XFEL Project Head Office, Sayo, Hyogo
679-5148, Japan}%

\author{T.~Ishikawa}
\affiliation{RIKEN, XFEL Project Head Office, Sayo, Hyogo
679-5148, Japan}%

\begin{abstract}
Phase-shift differences and amplitude ratios of the outgoing $s$ and $d$ 
continuum wave packets generated by two-photon ionization of helium atoms are 
determined from the photoelectron angular distributions obtained using velocity 
map imaging. Helium atoms are ionized with ultrashort extreme-ultraviolet free-electron 
laser pulses with a photon energy of 20.3, 21.3, 23.0, and 24.3 eV, produced 
by the SPring-8 Compact SASE Source test accelerator. The measured values of 
the phase-shift differences are distinct from scattering phase-shift 
differences when the photon energy is tuned to an excited level or Rydberg 
manifold.
The difference stems from the competition between resonant and non-resonant paths in two-photon ionization by ultrashort pulses.  Since the competition can be controlled in principle by the pulse shape, the 
present results illustrate a new way to tailor the continuum wave packet.
\end{abstract}

\date{\today}
\maketitle

Two-photon processes are well-known phenomena
and have been extensively investigated for decades 
both experimentally and theoretically. 
Also these processes have been used in a variety of applications in laser 
optics and spectroscopy. 
It is well known that the two-photon photoelectron angular 
distributions are directly related to the relative amplitudes and the relative phase between 
different partial waves~\cite{Dixit,Smith,Lambropoulos,Wang2001PRL,Reid,Kabachnik}.
However, these earlier works dealt with the laser pulses in the optical range 
whose pulse width is very long in comparison with the modern standard of femosecond laser technology. 
 The advent of extreme ultraviolet 
(EUV)~\cite{FLASH,SCSS} and x-ray~\cite{LCLS,SACLA}
free-electron lasers (FELs), with femtosecond pulse widths, 
has led to renewed interest in two-photon
processes in the EUV to x-ray regimes (see, e.g., ~\cite{Nagasono2007PRA,Varma2009PRA,Santra2009PRL,Young2010Nature,Cryan2010PRL,Fang2010PRL,Berrah2011PNAS,Doumy2011PRL,Salen2012PRL}).   
In the present Letter we address a new opportunity opened 
by the ultrashort EUV FEL pulses to deviate the phase shift difference between ionization channels from the scattering phase shift difference, which is otherwise intrinsic to the target atom or molecule. This will eventually open a new avenue to the coherent control of the continuum wave packets. (In this connection, see Ref.\ \cite{Dudovich2001PRL} for the control of the resonant two-color two-photon excitation yield and  Ref\ \cite{Wollenhaupt2009APB} for the control of the photoelectron angular distributions of the nonperturbative resonant multi-photon ionization with ultrashort polarization-shaped pulses. See also a very recent review article for photoelectron angular distributions~\cite{Reid2012MP}.)

The simplest possible two-photon process may be (single-color) 
two-photon single ionization of helium atoms.  
For theoretical study, see, for example, the work by 
Nikolopoulos \emph{et al.}~\cite{Nikolopoulos}, van der Hart and Bingham~\cite{Hart}, and references cited therein.  
Kobayashi  \emph{et al.}~\cite{Kobayashi1998OL} were the first to observe this process 
and used it for an autocorrelation measurement of high-order harmonic pulses, and
Moshammer \emph{et al.}~\cite{Moshammer2011OE} recently used it for an autocorrelation 
measurement of the EUV FEL pulses 
provided by the SPring-8 Compact SASE Source (SCSS) test accelerator~\cite{SCSS}.
The absolute two-photon ionization cross sections of He were measured using an intense 
high harmonic source~\cite{Hasegawa2005PRA} as well as 
the SCSS test accelerator~\cite{Sato2011JPB}.  
Hishikawa \emph{et al.}~\cite{Hishikawa2011PRL} recently investigated 
two- and three-photon ionization of He at the SCSS test accelerator 
by photoelectron spectroscopy using a magnetic bottle spectrometer.
The photoelectron angular distribution (PAD) for single-color two-photon ionization of helium, however, has not been investigated so far, though those from two-color two-photon (EUV + infrared) above-threshold ionization have recently been reported by Haber {\it et al.}~\cite{Haber2011PRA}.

Two-photon single ionization of helium produces a continuum electron wave packet 
which is a superposition of $s$ and $d$ partial waves.
The photoelectron angular distribution provides information about 
the ratio of amplitudes for the $s$ and $d$ partial waves and their relative
phase.  Extracting a phase shift difference from the PAD
that is observed from a photoexcited state is a well-established method. For
example, Haber \emph{et al.}~\cite{Haber2009PRA} excited the ground state
helium atom to the $1snp$ Rydberg states by high-order
harmonics and measured the PAD emitted from these excited states using IR and
UV lasers as ionizing pulses.  A
similar experiment was also performed by O'Keeffe \emph{et
al.}~\cite{Okeeffe2010JPCS} using synchrotron radiation for the excitation
and a laboratory laser for the ionization. 
Both these experiments confirmed that the relative phase extracted from measured PADs 
resulting from sequential two-color excitation and ionization agrees well 
with the theoretically predicted scattering phase shift difference. 
In contrast, as theoretically 
predicted by Ishikawa and Ueda~\cite{Ishikawa2012PRL}, the situation is significantly different 
for two-photon ionization by an intense short pulse as a competition between resonant and 
non-resonant ionization paths leads to a relative phase between $s$ and $d$ that is distinct 
from the corresponding scattering phase difference. It is expected that this change 
in the phase difference can be revealed by means of a PAD measurement.
 
In the present study, we use velocity map imaging (VMI)~\cite{Parker,Rouzee2011PRA}
to measure the PAD from two-photon ionization 
of He by intense, femtosecond EUV FEL pulses. The 
anisotropy parameters are obtained from the PAD to extract the phase differences $\delta$ and the amplitude ratios of the $s$ and $d$ partial waves at four different photon energies ($\hbar\omega = 20.3, 21.3, 23.0$, and 24.3 eV). Our results show the presence of an extra phase shift due to a competition between resonant and non-resonant paths, in agreement with our recent theoretical prediction~\cite{Ishikawa2012PRL} and simulation results obtained by solving the full time-dependent Schr\"odinger equation.  

The experiments were carried out with the SCSS test accelerator in Japan. 
This FEL light source provided EUV pulses with a duration of 
$\sim 30$ fs~\cite{Moshammer2011OE} and a full-width-at-half-maximum 
spectral width of $\sim 0.2$ eV \cite{Hikosaka2010PRL}.
The photon energies were selected to be 20.3 eV, 21.3 eV , 23.0 eV, and 24.3 eV. 
A photon energy of 20.3 eV is well below the excitation energy to the lowest
resonance $1s2p~^1P$ (21.218 eV)~\cite{NIST} and, thus, ionization is expected to
be dominated by direct, non-resonant two-photon ionization. 
Photon energies 21.3 and 23.0 eV are close to the $1s2p~^1P$ 
and $1s3p~^1P$ (23.087 eV)~\cite{NIST}
resonances. According to theoretical predictions~\cite{Ishikawa2012PRL}, 
we may expect that competition between resonant and non-resonant paths 
can be seen.
A photon energy of 24.3 eV corresponds to excitation to the $1snp~^1P$ 
Rydberg manifold ($n \sim 7$).  The spectral width covers several
Rydberg members from $1s6p~^1P$ to $1s9p~^1P$. In this condition
we also expect that resonant and non-resonant ionization compete.

The FEL beam from the SCSS test accelerator was steered by two upstream 
plane SiC mirrors, passed a gas monitor detector (GMD), and then entered 
the prefocusing system of the beam line. 
The GMD was calibrated using a cryogenic radiometer~\cite{Kato}. 
The average pulse energy measured by the
GMD during the experiments was $7-11~\mu$J, 
with a standard deviation of $2-4$~$\mu$J. 
The focusing system,  with a focal length of 1 m, consisted of a pair 
of elliptical and cylindrical mirrors coated with SiC~\cite{Ohashi}. The  
reflectivity of each mirror was 70 \%.  Before entering the
interaction chamber, the FEL beam passed through two sets of light
baffles, each consisting of three skimmers with 4.0 mm and 3.5 mm
diameters, respectively. These baffles successfully removed the
majority of the scattered light specularly and non-specularly
reflected by the two mirrors, without reducing the photon flux.   
The FEL beam was then focused on a helium beam at the center 
of a VMI spectrometer~\cite{Gryzlova2011PRA}. 
The measured focal spot size was $\sim 13 \mu$m in
radius, resulting in an average intensity of typically 
$2-3 \times 10^{13}$ W/cm$^{2}$.

\begin{figure}
\includegraphics[width=80mm]{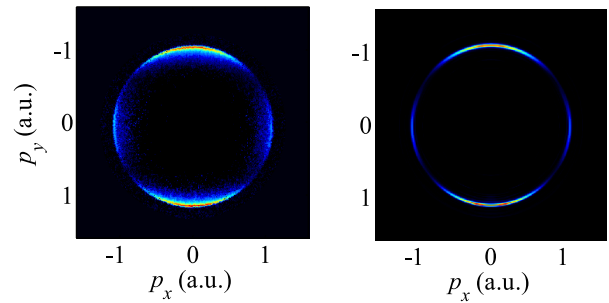}
\caption{Raw (left) and inverted (right) photoelectron images 
from two-photon ionization of helium at 20.3 eV photon energy.}
 \label{fig:image}
\end{figure}

Electrons produced by two-photon ionization of the helium atoms 
by the FEL pulses were accelerated, perpendicularly to both the
propagation and linear polarization axes of the FEL beam, 
towards a position-sensitive detector
consisting of a set of microchannel plates (MCPs) followed by a
phosphor screen. The positions of detected electrons were 
recorded using a gated CCD camera synchronized to the arrival 
of the FEL pulse in the interaction chamber. 
A 200~ns electrical gate pulse was applied to the back of
the MCPs.  
The photoelectron angular distribution (PAD) has cylindrical 
symmetry along the FEL polarization, and we can retrieve the 
three-dimensional (3D) photoelectron momentum distribution 
from the raw two-dimensional (2D) image
using a mathematical inversion procedure, 
which for each radial momentum leads to an expression of 
the photoelectron angular distribution in terms of Legendre coefficients (see Eq. (1) below).
Examples of raw and inverted images are given in Fig.\ \ref{fig:image}.

\begin{figure}
\includegraphics[width=8cm]{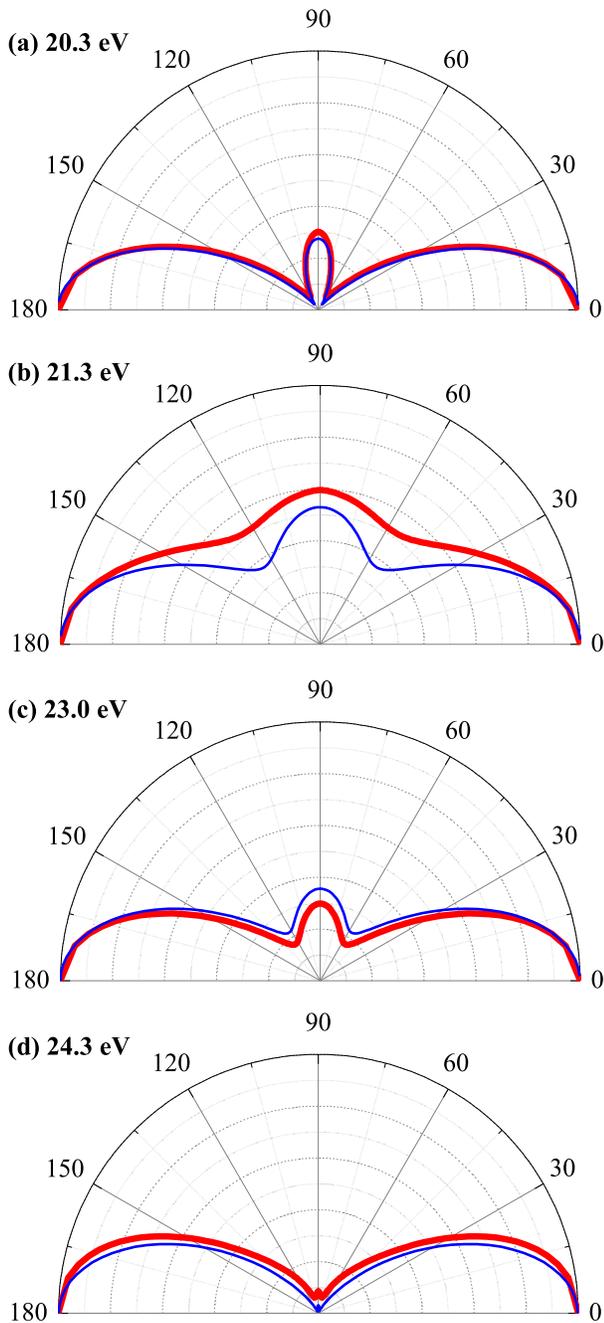}
\caption{(Color online) Measured (thick red lines) and calculated (thin blue lines) photoelectron angular
distributions for two-photon ionization of helium at photon energies
of 20.3, 21.3, 23.0, and 24.3 eV.} \label{fig:PAD}
\end{figure}

\begin{table}
  \centering 
  \caption{Experimentally obtained anisotropy parameters $\beta_2$ and $\beta_4$, and the extracted values of $W$ and $\Delta$.}
  \label{table:table01}
  \begin{tabular}{ccccc}
\hline
$\hbar\omega$ (eV) & $\beta_2$ & $\beta_4$ & $W$ & $\Delta$ \\
\hline
20.3 & 1.14$\pm$0.07 & 1.96$\pm$0.03 & 0.561$\pm$0.016 & 1.60$\pm$0.05 \\
21.3 & 0.268$\pm$0.019 & 0.384$\pm$0.063 & 2.39$\pm$0.23 & 1.61$\pm$0.04 \\
23.0 & 0.948$\pm$0.010 & 1.32$\pm$0.15 & 0.977$\pm$0.116 & 1.67$\pm$0.04 \\
24.3 & 2.11$\pm$0.10 & 0.841$\pm$0.006 & 1.43$\pm$0.01 & 2.47$\pm$0.07 \\
\hline
\end{tabular}
\end{table}

Figure \ref{fig:PAD} {displays the PADs $I(\cos\theta)$ obtained 
at four different photon energies,  
as a function of cosine of the polar angle $\theta$ relative 
to the polarization axis.  
The PADs $I(\cos\theta)$  can be described by the following expression:
\begin{equation}\label{Legendre}
I(\cos\theta) = \frac{I_0}{4\pi} \left[ 1 + \beta_2 P_2(\cos\theta)+ \beta_4 P_4(\cos\theta) \right] \;,
\end{equation}
where $I_0$ is the angle-integrated intensity, 
{$\beta_2$ and $\beta_4$ are the anisotropy parameters associated 
with the second- and fourth-order Legendre polynomials $P_2(x)$ and $P_4(x)$, respectively. 
Values of $\beta_2$ and $\beta_4$ obtained from the experimental PADs are listed in Table 1.

To investigate the processes involved in the two photon ionization of He, we have performed numerical simulations, by solving the full-dimensional two-electron time-dependent Schr\"odinger equation (TDSE) using the time-dependent close-coupling method \cite{ATDI2005,Parker2001JPB,Pindzola1998PRA,Pindzola1998JPB,Colgan2001JPB}. We have employed chaotic pulses with a mean intensity of $2.5\times 10^{13}\,{\rm W/cm}^2$, generated by the partial-coherence method described in Ref.\ \cite{Pfeifer2010OL}, for a coherence time of 8 fs and a mean pulse width of 28 fs (full width at half maximum), as recently measured by second-order autocorrelation~\cite{Moshammer2011OE}, both assumed to have Gaussian profiles on average.
One can see a good agreement between the experimental and simulation results in
Fig.\ \ref{fig:PAD} as well as in Fig.\ \ref{fig:phaseshift} below.

The PAD results from an interference of the $s$ and $d$ partial waves, and can be expressed as,
\begin{eqnarray}
&\propto& \left| c_{0}e^{i\delta_{sc,0}}Y_{00}-c_2e^{i\delta_{sc,2}}Y_{20}\right|^2 \nonumber\\
&=& \left| |c_0|e^{i\delta_0}Y_{00}-|c_2|e^{i\delta_2}Y_{20}\right|^2,
\end{eqnarray}
where $c_l$ denotes the complex amplitude of a final state with an angular momentum $l$, $\delta_{sc,l}$ the scattering phase shift intrinsic to the corresponding continuum eigen function, and $\delta_l = \arg c_l + \delta_{sc,l}$ the phase of each partial wave.
If we define $W=|c_0/c_2|$ and $\Delta=\delta_0-\delta_2$, then, these are related to the anisotropy parameters as,
\begin{equation}
\label{eq:beta2and4}
\beta_2=\frac{10}{W^2+1}\left[\frac{1}{7}-\frac{W}{\sqrt{5}}\cos\Delta\right], \quad
\beta_4=\frac{18}{7(W^2+1)},
\end{equation}
and thus $W$ and $\delta$ can be extracted from the PAD. The experimentally obtained values of $W$ and $\Delta$ are
listed in Table I. Furthermore, the experimental values of $W$ and $\Delta$ are compared with values extracted from TDSE simulations
in Fig.\ \ref{fig:phaseshift} as a function of photon energy $\hbar\omega$.
The agreement between experimental and theoretical values are reasonable for both $W$ 
and $\Delta$. For comparison, theoretical values of the scattering
phase shift difference $\Delta_{sc}\equiv\delta_{sc,0}-\delta_{sc,2}$~\cite{Gien2002JPB} 
are also plotted by the solid line.

Within the framework of the second-order time-dependent perturbation theory
\cite{Ishikawa2012PRL}, $c_l$ can be defined in such a way that its real and imaginary 
parts correspond to the resonant and non-resonant paths, respectively. 
If the pulse is non-resonant, $c_0$ and $c_2$ are pure imaginary, resulting in $\Delta=\Delta_{sc}$. 
In the present study, the measurement at $\hbar\omega = 20.3\, {\rm eV}$ 
corresponds to this situation; indeed, we find $\Delta \approx \Delta_{sc}$
in this case in Fig.\ \ref{fig:phaseshift}.

Let us now turn to the situation where the pulse is resonant with an excited state or Rydberg manifold.
If a resonant two-photon ionization path is dominant, $\Delta$ is again close to $\Delta_{sc}$, 
since $c_0$ and $c_2$ are both real. 
On the other hand, if the contributions from both the resonant via a single or several resonant levels)
and non-resonant paths (via all the intermediate levels) 
are present, then $\Delta\ne\Delta_{sc}$ in general \cite{Ishikawa2012PRL} and 
an extra phase shift difference $\Delta_{ex}\equiv\Delta - \Delta_{sc}=\arg c_0/c_2$ 
occurs that can be viewed as a measure of the competition between them.
In the present study, the pulses with $\hbar\omega=21.3, 23.0$, 
and $24.3$ eV induce resonant two-photon ionization 
via $1s2p\,^1P$, $1s3p\,^1P$, and a Rydberg manifold ($1snp\,^1P$ with $n=6-9$), 
respectively. 
We can see in Fig.\ \ref{fig:phaseshift} that the relative phase $\Delta$ deviates 
from the scattering phase shift difference $\Delta_{sc}$ for these three photon energies; 
the difference $\Delta_{ex}$
increases gradually with increasing photon energy and becomes very significant at 24.3 eV. 
At $\hbar\omega=21.3$ and $23.0$ eV, the simulation values of $\Delta_{ex}$ are slightly smaller
than for lower intensity, indicating the departure from the perturbative limit due to the high intensity.
The observed deviation $\Delta_{ex}$
clearly demonstrates the presence of a competition between resonant and non-resonant paths 
in the present experiments, as recently predicted in Ref.\ \cite{Ishikawa2012PRL}. 
This situation presents a contrast to the case of the photoionization from excited $p$ 
states \cite{Haber2009PRA,Okeeffe2010JPCS}, 
where the non-resonant path is absent and, as a result, $\Delta=\Delta_{sc}$. 
Although the competition has been implicitly used in coherent control of resonance-enhanced 
multi-photon processes (see, e.g., \cite{Dudovich2001PRL,Wollenhaupt2009APB}), intermediate levels other than the resonant level are neglected in most cases.
In the present study, on the other hand, the contribution from non-resonant intermediate levels is essential
to account for $\Delta_{ex}$ \cite{Ishikawa2012PRL}, which explains why $\Delta_{ex}$ is larger for a higher photon energy, i.e., 
for smaller level spacing.

\begin{figure}[tb]
\centering
\includegraphics[width=80mm]{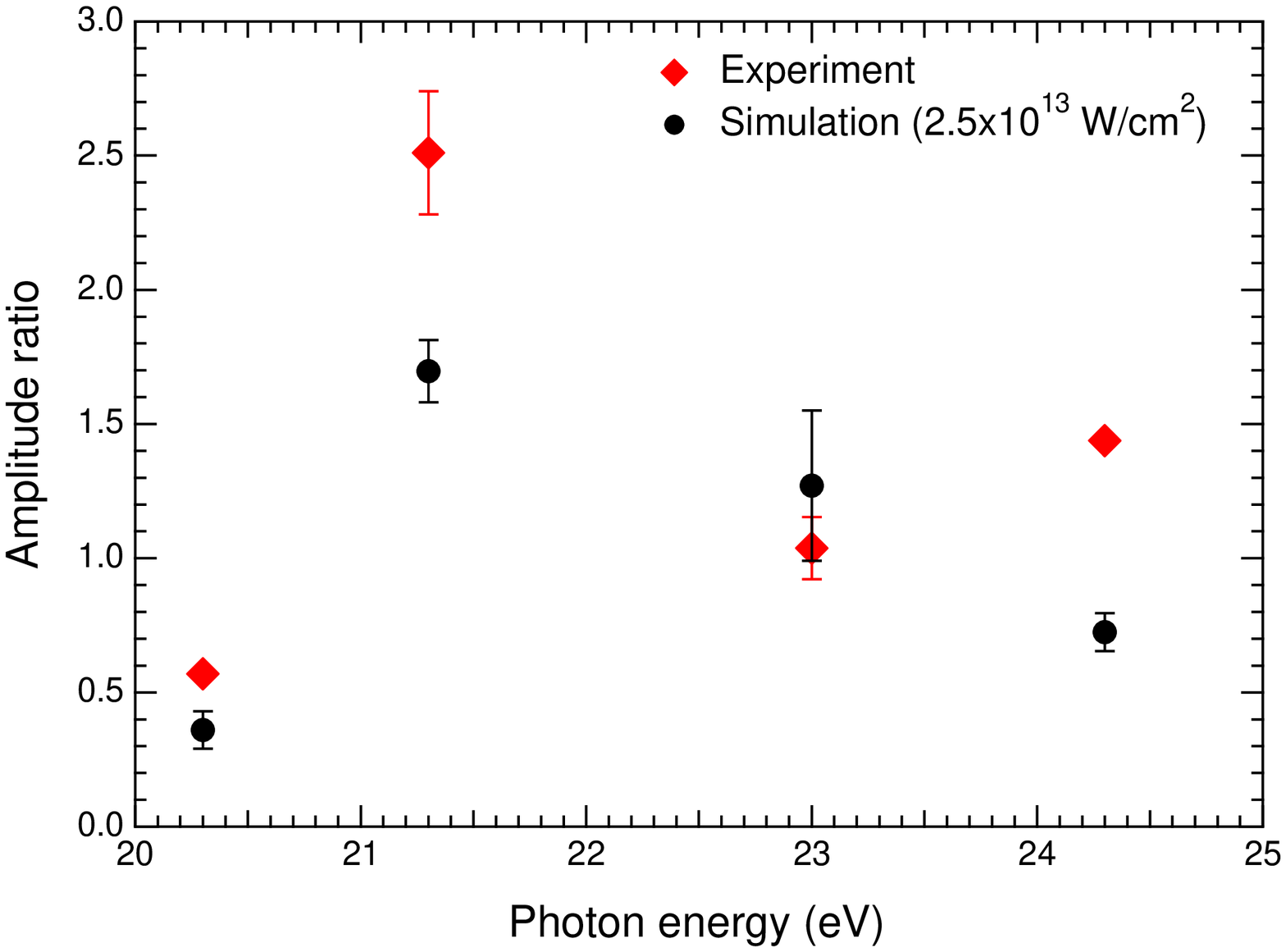}
\includegraphics[width=80mm]{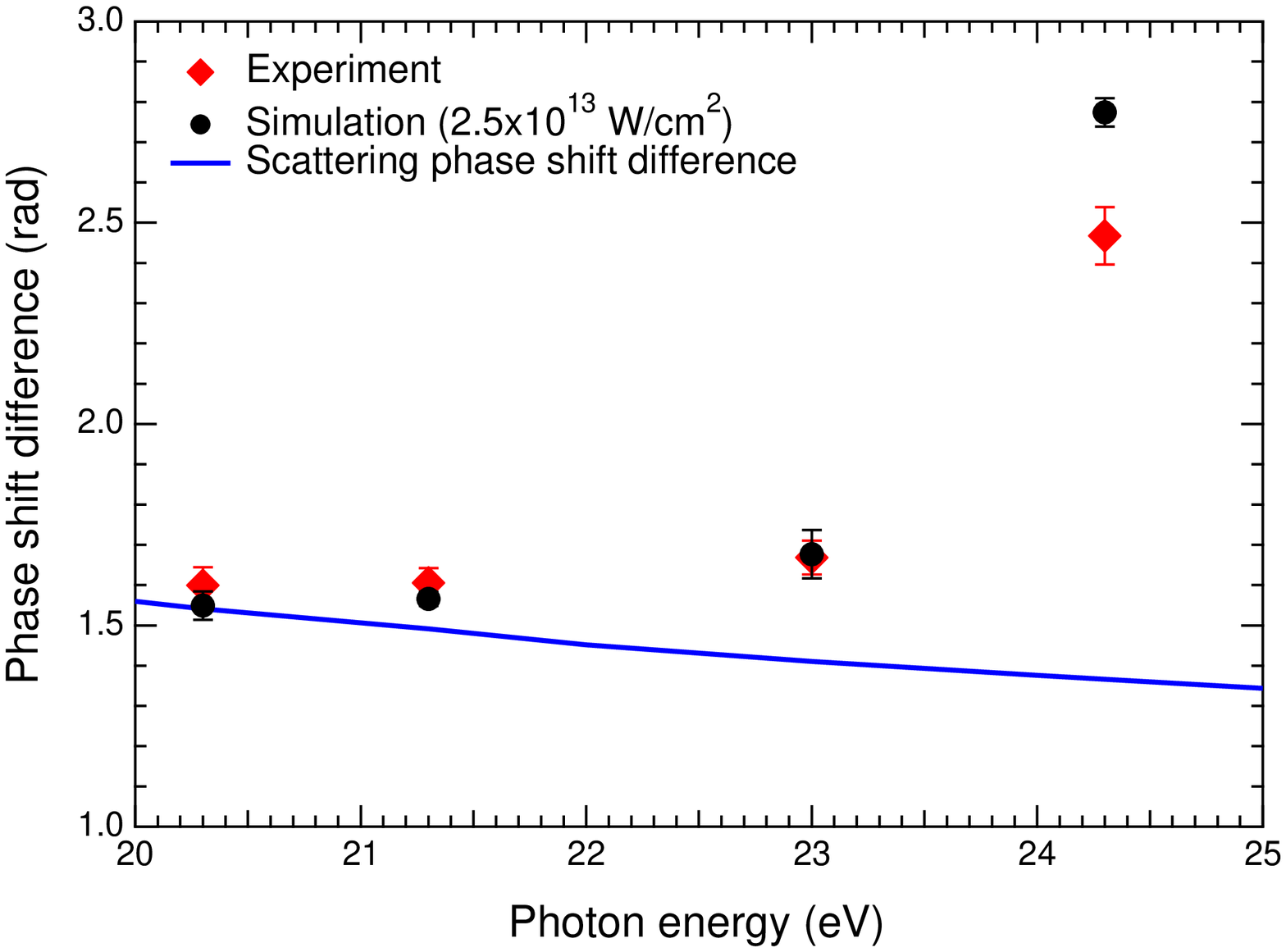}
\caption{(Color online) Amplitude ratio $W$ (upper panel) and phase shift difference 
(relative phase) $\Delta$ (lower panel) extracted from
experimental and theoretical PADs. Theoretical scattering phase shift
difference $\Delta_{sc}$ \cite{Gien2002JPB} is also included in the lower panel.}
\label{fig:phaseshift}
\end{figure}

It may be worth pointing out the similarity between two-photon ionization via a 
Rydberg manifold and two-photon above-threshold ionization.
In the case of the 24.3 eV excitation, the intense ultrashort EUV pulses used in the
present study coherently excite several Rydberg states $1snp\,^1P$ with $n=6-9$. 
In such a situation, the Rydberg manifold
behaves similarly to the continuum 
near the threshold and both the relative phase $\Delta$~\cite{Ishikawa2012PRL} 
and the TPI yield~\cite{IshikawaJMO2010} would smoothly vary when measured by
increasing the photon energy across the ionization threshold.
It should be noted that the extra phase shift difference due to free-free transitions 
plays a significant role in recently observed time delays in photoemission 
by attosecond EUV pulses \cite{Schultze2010Science,Kluender2011PRL}.

In conclusion, we have measured the PAD
from two-photon ionization of He by intense, femtosecond EUV FEL pulses
provided by the SCSS test accelerator in Japan using a VMI spectrometer. 
From the anisotropy parameters of the PAD, we extracted phase-shift differences $\Delta$ 
and amplitude ratios $W$ of the $s$ and $d$ partial waves at four different photon energies 
($\hbar\omega = 20.3, 21.3, 23.0$, and 24.3 eV). 
As a result, we have demonstrated that competition between resonant 
and non-resonant processes in two-photon ionization
by intense femtosecond pulses causes an additional phase shift 
in the photoelectron wavepacket.
The competition can in principle be controlled by chirping the EUV pulses,
which may pave a way to tailor continuum wave packets.
Such an experiment will become feasible in the near future at FEL facilities
where the pulses can be controlled in the range of a few to
a few tens of fs.

We are grateful to the SCSS Test Accelerator Operation Group at
RIKEN for continuous support in the course of the studies and to
the staff of the technical service section in IMRAM, Tohoku
University, for their assistance in constructing the apparatus.
This study was supported by the X-ray Free Electron Laser
Utilization Research Project of the Ministry of Education,
Culture, Sports, Science and Technology of Japan (MEXT), 
by the Management Expenses Grants for National Universities 
Corporations from MEXT, by Grants-in-Aid for Scientific Research 
from JSPS (No. 21244062 and  No. 22740264), and IMRAM research program.
K.L.I. gratefully acknowledges support by the APSA Project (Japan), KAKENHI (No. 23656043 and No. 23104708), the Project of Knowledge Innovation Program (PKIP) of Chinese Academy of Sciences (Project No. KJCX2.YW.W10),
and the Cooperative Research Program of Network Joint Research Center for
Materials and Devices. The research of A.H., A.R. and M.V. is  part of the research program of the ``Stichting voor Fundamenteel Onderzoek der Materie (FOM)'', which is financially supported by the ``Nederlandse organisatie voor Wetenschappelijk Onderzoek
(NWO)''.  P.J. acknowledges support from the Swedish Research Council and the Swedish Foundation for Strategic Research.


\begin{references}
%

\bibitem{Dixit}
S.N. Dixit and P. Lambropoulos, {\bf 27}, 861 (1983). 

\bibitem{Smith}
S.J. Smith, G. Leuchs Adv. At. Mol. Phys. {\bf 24}, 157-221 (1988).

\bibitem{Lambropoulos}
P. Lambropoulos, P. Maragikis, and J. Zhang, 
Phys. Rep. {\bf 305}, 203 (1998).

\bibitem{Wang2001PRL}
Z.-M. Wang and D.S. Elliott, \prl {\bf 87}, 173001 (2001).

\bibitem{Reid}
K.L. Reid, Annu. Rev. Chem. {\bf 54}, 397 (2003).

\bibitem{Kabachnik}
N.M. Kabachnik, S. Fritzsche, A.N. Grum-Grzhimailo, M. Meyer, and K. Ueda, 
Phys. Rep. {\bf 451}, 155 (2007).

\bibitem{FLASH}
W. Ackermann  {\it et al.}, Nat. Photonics, {\bf 1}, 336 (2007).

\bibitem{SCSS}
T. Shintake {\it et al.}, Nat. Photonics, {\bf 2}, 555 (2008).

\bibitem{LCLS}
P. Emma  {\it et al.}, Nat. Photonics, {\bf 4}, 641 (2010).

\bibitem{SACLA}
T. Ishikawa {\it et al}., Nat. Photonics \textbf{6}, 540 (2012).

\bibitem{Nagasono2007PRA}
M. Nagasono  {\it et al.}, 
Phys. Rev. A {\bf75}, 051406 (2007).

\bibitem{Varma2009PRA}
H. R. Varma, M. F. Ciappina, N. Rohringer, and R. Santra, Phys. Rev. A {\bf 80}, 053424 (2009). 

\bibitem{Santra2009PRL}
R. Santra, N. V. Kryzhevoi, and L. S. Cederbaum, 
Phys. Rev. Lett. {\bf 103}, 013002 (2009).

\bibitem{Young2010Nature}
L. Young {\it et al.},  Nature (London) {\bf 466}, 56 (2010).

\bibitem{Cryan2010PRL}
J..P. Cryan  {\it et al.}, Phys. Rev. Lett. {\bf 105}, 083004 (2010).

\bibitem{Fang2010PRL}
L. Fang  {\it et al.}, Phys. Rev. Lett. {\bf 105}, 083005 (2010).

\bibitem{Berrah2011PNAS}
N. Berrah  {\it et al.}, Proc. Nat. Acad. Sci. {\bf 108}, 16912 (2011).

\bibitem{Doumy2011PRL}
G. Doumy  {\it et al.},  Phys. Rev. Lett. {\bf 106}, 083002 (2011).

\bibitem{Salen2012PRL}
P. Sal{\'e}n  {\it et al.}, Phys. Rev. Lett. {\bf 108}, 153003 (2012).

\bibitem{Dudovich2001PRL} N. Dudovich, B. Dayan, S.M. Gallagher Faeder, and Y. Silberberg, \prl {\bf 86}, 47 (2001).

\bibitem{Wollenhaupt2009APB}
M. Wollenhapt M. Klug, J. K{\"o}hler, T. Bayer, C. Sarpe-Tudoran, and T. Baumert, 
Appl. Phys. B {\bf 95}, 245 (2009).

\bibitem{Reid2012MP}
K.L. Reid, Mol. Phys. {\bf 110}, 131 (2012).

\bibitem{Nikolopoulos} 
L.A.A. Nikolopoulos and P. Lambropoulos, J. Phys. B: At. Mol. Opt. Phys. {\bf 34}, 545 (2001). 

\bibitem{Hart}
H.W. van der Hart and P. Bingham, J. Phys. B: At. Mol. Opt. Phys. {\bf 38}, 207 (2005).

\bibitem{Kobayashi1998OL}
Y. Kobayashi, T. Sekikawa, Y. Nabekawa, and S. Watanabe, Opt. Lett. {\bf 23}, 64 (1998).

\bibitem{Moshammer2011OE} 
R. Moshammer {\it et al.}, Opt. Express {\bf 19}, 21698 (2011).

\bibitem{Hasegawa2005PRA} 
H. Hasegawa, E. J. Takahashi, Y. Nabekawa, K. L. Ishikawa, and K. Midorikawa, Phys. Rev. A {\bf 71}, 023407 (2005).

\bibitem{Sato2011JPB}
T. Sato  {\it et al.}, J. Phys. B: At. Mol. Opt. Phys. {\bf 44}, 161001 (2011).

\bibitem{Hishikawa2011PRL}
A. Hishikawa  {\it et al.},  Phys. Rev. Lett. {\bf 107}, 243003 (2011).

\bibitem{Haber2011PRA} 
L. H. Haber, B. Doughty, and S. R. Leone, \pra {\bf 84}, 013416 (2011).

\bibitem{Haber2009PRA} 
L. H. Haber, B. Doughty, and S. R. Leone, \pra {\bf 79}, 031401(R) (2009).

\bibitem{Okeeffe2010JPCS}
P. O'Keeffe P. Bolognesi, R. Richter, A. Moise, E. Ovcharenko, L. Pravica, R. Sergo, L.
Stebel, G. Cautero, and L. Avaldi1, J. Phys.: Conf. Ser. {\bf 235},  012006 (2010).

\bibitem{Ishikawa2012PRL}
K.L. Ishikawa and K. Ueda, Phys. Rev. Lett. {\bf 108}, 033003 (2012).

\bibitem{Parker}
A. T. J. B. Eppink and D. H. Parker, Rev. Sci. Instrum. {\bf 68}, 3477 (1997).

\bibitem{Rouzee2011PRA}
A. Rouzee  {\it et al.}, 
Phys. Rev. A {\bf  83}, 031401(R) (2011).

\bibitem{Hikosaka2010PRL}
Y. Hikosaka \emph{et al.}, Phys. Rev. Lett. \textbf{105}, 133001
(2010).

\bibitem{NIST} http://physics.nist.gov/PhysRefData/ASD/levels\_form.html.

\bibitem{Kato}
M. Kato \emph{et al.},  Nucl. Instrum. Methods Phys. Res. A
\textbf{612}, 209 (2009).

\bibitem{Ohashi}
H. Ohashi {\it et al.}, Nucl. Instrum. Methods Phys. Res., Sect. A
\textbf{649}, 58 (2011).

\bibitem{Gryzlova2011PRA}
E.V. Gryzlova   {\it et al.}, 
Phys. Rev. A {\bf  84}, 063405 (2011).

\bibitem{Pindzola1998PRA} M. S. Pindzola and F. Robicheaux, Phys. Rev. A {\bf 57}, 318 (1998).

\bibitem{Pindzola1998JPB} M. S. Pindzola and F. Robicheaux, J. Phys. B {\bf 31}, L823 (1998).

\bibitem{Colgan2001JPB} J. Colgan, M. S. Pindzola, and F. Robicheaux, J. Phys. B {\bf 34}, L457 (2001).

\bibitem{Parker2001JPB} J. S. Parker, L. R. Moore, K. J. Meharg, D. Dundas, and K. T. Taylor, J. Phys. B {\bf 34}, L69 (2001).

\bibitem{ATDI2005} K. L. Ishikawa and K. Midorikawa, Phys. Rev. A {\bf 72}, 013407 (2005).

\bibitem{Pfeifer2010OL} T. Pfeifer, Y. Jiang, S. D\"usterer, R. Moshammer, and J. Ullrich, Opt. Lett. {\bf 35}, 3441 (2010).

\bibitem{Gien2002JPB} T. T. Gien, J. Phys. B {\bf 35}, 4475 (2002).


\bibitem{IshikawaJMO2010}
K. L. Ishikawa, Y. Kawazura, and K. Ueda, J. Mod. Opt. {\bf 57}, 999 (2010).

\bibitem{Schultze2010Science} M. Schultze {\it et al.}, Science {\bf 328}, 1658 (2010).

%
\bibitem{Kluender2011PRL} K. Kl\"under {\it et al.}, Phys. Rev. Lett. {\bf 106}, 143002 (2011).

%
%
%
%


\end{references}
\end{document}